# Simultaneous weak measurement of non-commuting observables

M. A. Ochoa[1], W. Belzig[2] and A. Nitzan[1,3]

[1]Department of Chemistry, University of Pennsylvania, Philadelphia, PA 9104

[2]Depatment of Physics, University of Konstanz, D-78457 Konstanz, Germany

[3] School of Chemistry, Tel Aviv University, Tel Aviv 69978, Israel

## Abstract

In contrast to a projective quantum measurement in which the system is projected onto an eigenstate of the measured operator, in a weak measurement the system is only weakly perturbed while only partial information on the measured observable is obtained. A full description of such measurement should describe the measurement protocol and provide an explicit form of the measurement operator that transform the quantum state to its post measurement form. A simultaneous measurement of non-commuting observables cannot be projective, however the strongest possible such measurement can be defined as providing their values at the smallest uncertainty limit. Starting with the Arthurs and Kelly (AK) protocol for such measurement of position and momentum, we derive a systematic extension to a corresponding weak measurement along three steps: First, a plausible form of the weak measurement operator analogous to the Gaussian Kraus operator often used to model a weak measurement of a single observable is obtained by projecting a naïve extension (valid for commuting observable) onto the corresponding Gabor space. Second, we show that the so obtained set of measurement operators satisfies the normalization condition for the probability to obtain given values of the position and momentum in the weak measurement operation, namely that this set constitutes a positive operator valued measure (POVM) in the position-momentum space. Finally, we show that the so-obtained measurement operator corresponds to a generalization of the AK measurement protocol in which the initial detector wavefunctions is suitable broadened.

# Introduction

The possibility of and limitations on simultaneous measurement of non-commuting variables has repeatedly attracted attention of many theorists over the last half century,[1-11] and is attracting renewed attention recently as new techniques for such measurements are manifested.[12, 13] Arthurs and Kelly (AK)[1] have generalized the von-Neumann's concept of quantum measurement[14] to describe such simultaneous measurement of position and momentum of a quantum particle by coupling it to two mutually independent detectors set to detect these variables. Obviously, such a measurement cannot determine position and momentum exactly. As explained in Appendix A, we find it useful to present these results in a form which is physically different from, but mathematically equivalent to, that of Ref. [1]. In this form, the interaction between the measured particle and the detectors

$$H_{int} = K(\hat{p}_1 \hat{x} + \hat{x}_2 \hat{p}), \tag{1}$$

is set so as to shift the position of detector 1 and the momentum of detector 2 by amounts that correspond to the particle's position and momentum. The highest possible accuracy is obtained when the interaction is set to operate between time 0 and $K^{-1}$ and is assumed to dominate the evolution during this time, and when the initial wavefunctions of the measured quantum particle and the two detectors are respectively $\Psi_B(\bar{x})$ and[15]

$$D_1(\bar{x}_1) = \left(\frac{2}{\pi b}\right)^{1/4} e^{-\bar{x}_1^2/b} \quad ; \quad \tilde{D}_2(\bar{p}_2) = \left(\frac{2b}{\pi}\right)^{1/4} e^{-b\bar{p}_2^2} \tag{2}$$

(the wavefunction of detector 2 is expressed in the momentum representation and $\hbar$ is taken 1 throughout) where $b$ is an arbitrary parameter of dimension [length$^2$]. Two main results obtained by AK are expressed as follows: First, if projective measurement of the detector states determine the particle position and momentum to be $\bar{x}_m$ and $\bar{p}_m$, then the normalized wavefunction after the measurement is given up to a phase by (see Appendix A)

$$\Psi_A(\bar{x}) = \left(\frac{1}{\pi b}\right)^{1/4} e^{i\bar{p}_m \bar{x}} e^{-\frac{1}{2b}(\bar{x}-\bar{x}_m)^2} \tag{3}$$

Irrespective of the form of $\Psi_B$. Second, the joint probability density to find the values $\bar{x}_m$ and $\bar{p}_m$ is

$$P(\bar{x}_m, \bar{p}_m) = \frac{1}{2\sqrt{\pi^3 b}} \left| \int d\bar{x} \Psi_B(\bar{x}) e^{-\frac{(\bar{x}-\bar{x}_m)^2}{2b}} e^{-i\bar{x}\bar{p}_m} \right|^2 \qquad (4)$$

It is important to note that the choice made by AK to set $\tau \equiv Kt = 1$ implies that the values obtained for the detectors variables $\bar{x}_m$ and $\bar{p}_m$ reflect the values of the position and momentum, $\bar{x}$ and $\bar{p}$, of the system itself. This may be contrasted with shorter time measurements, see Eq. (34) below.

As noted by later authors, see, e.g. [2,3] the post-measurement wavefunction essentially represents a coherent state. To set the formal relationship we transform to dimensionless variables according to

$$\bar{x}/\sqrt{2b} \to x, \quad \bar{p}\sqrt{b/2} \to p, \quad \Psi(\bar{x}) \to (2b)^{1/4} \Psi(x) \qquad (5)$$

so that

$$\Psi_A(x) \to \left(\frac{2}{\pi}\right)^{1/4} e^{2ip_m x} e^{-(x-x_m)^2}; \quad P(x_m, p_m) = \sqrt{\frac{2}{\pi^3}} \left| \int dx \Psi_B(x) e^{-(x-x_m)^2} e^{-2ixp_m} \right|^2 \qquad (6)$$

We now define the vector $\alpha_m$ and the coherent state $|\alpha_m\rangle$ by

$$\alpha_m = (x_m, p_m), \qquad (7)$$

(sometimes conveniently represented as a complex number $\alpha_m = \alpha(x_m, p_m) = x_m + ip_m$), and

$$\langle x | \alpha_m \rangle = \left(\frac{2}{\pi}\right)^{1/4} e^{2ip_m x} e^{-(x-x_m)^2} \qquad (8)$$

and note that these coherent states are normalized, $\langle \alpha | \alpha \rangle = \int_{-\infty}^{\infty} dx |\langle x | \alpha \rangle|^2 = 1$ and satisfy the standard closure relationship

$$\pi^{-1} \int d^2\alpha |\alpha\rangle\langle\alpha| = \pi^{-1} \int_{-\infty}^{\infty} dx_m \int_{-\infty}^{\infty} dp_m |\alpha\rangle\langle\alpha| = \hat{I} \quad \text{(unit operator).} \qquad (9)$$

We can then cast the maximum accuracy simultaneous measurement of position and momentum that yields the values $x_m$ and $p_m$ (rendered dimensionless as defined above, using an arbitrary parameter $b$ of dimension [length$^2$] that, as seen above can be related to the initial detectors' wavefunctions) in terms of the Kraus operator[16]

$$\hat{K}_{\alpha_m} = \frac{1}{\sqrt{\pi}} |\alpha_m\rangle\langle\alpha_m| \qquad (10)$$

which relates the wavefunction after the measurement to that before it by

$$\Psi_A = \hat{K}_{\alpha_m} \Psi_B \tag{11}$$

and, in view of (9), fulfills the completeness condition

$$\int d^2\alpha \hat{K}_\alpha \hat{K}_\alpha^\dagger = \hat{I} \quad . \tag{12}$$

Since $\hat{K}_\alpha \hat{K}_\alpha^\dagger$ are positive and satisfy (12) they constitute a positive operator valued measure (POVM).

The measurement described by the operator (10) is not strong in the usual von Neumann sense. Indeed, as shown in [1], Eq. (4) implies that the variances in the measured quantities $x_m$ and $p_m$ satisfy

$$\langle \delta \bar{x}_m^2 \rangle = \langle \delta \bar{x}^2 \rangle_B + b/2 \ ; \ \langle \delta \bar{p}_m^2 \rangle = \langle \delta \bar{p}^2 \rangle_B + 1/(2b) \tag{13}$$

(or $\langle \delta x_m^2 \rangle = \langle \delta x^2 \rangle_B + 1/4$, $\langle \delta p_m^2 \rangle = \langle \delta p^2 \rangle_B + 1/4$), where $\langle \delta \hat{O}^2 \rangle_B = \langle \Psi_B | \hat{O}^2 | \Psi_B \rangle - \langle \Psi_B | \hat{O} | \Psi_B \rangle^2$. The increased uncertainty is in turn manifested in the position and momentum variances associated with the final wavefunction $\Psi_A$, Eq. (3) or $\Psi_A = |\alpha_m\rangle$, Eq. (6). Still, this measurement is the strongest possible for the simultaneous determination of position and momentum in the sense that the results adhere to the minimum possible uncertainty.

The most accurate determination of a single observable A corresponds to a projective measurement that yields an eigenvalue $a_j$ of the Hermitian operator $\hat{A}$ and leaves the system in the corresponding eigenfunction $\phi_j$ with probability $|\langle \phi_j | \Psi_B \rangle|^2$ (again, $\Psi_B$ is the system wavefunction before measurement). Weaker measurements, which result in less drastic effects on the system state at the cost of yielding less information, can be modelled in many ways. A particularly convenient one is described by the Gaussian Kraus operator

$$\hat{K}_a^\lambda \equiv \left(\frac{2\lambda}{\pi}\right)^{1/4} \exp\left[-\lambda(a - \hat{A})^2\right] \tag{14}$$

where the real number $a$ is the result of the measurement and $\lambda$ represents its weakness. The operation of $\hat{K}_a^\lambda$ is most easily seen when expressing the initial wavefunction in the basis of eigenstates of $\hat{A}$, $\Psi_B = \sum_j c_j \phi_j$, with the corresponding eigenvalues $a_j$

$$\Psi_A = \hat{K}_a^\lambda \Psi_B = \left(\frac{2\lambda}{\pi}\right)^{1/4} \sum_j c_j \exp\left[-\lambda(a-a_j)^2\right] \phi_j \qquad (15)$$

When $\lambda \to 0$ all results are possible, that is no information is obtained and the (normalized) wavefunction remains intact. When $\lambda \to \infty$ only an eigenvalue can be obtained and the wavefunction is projected onto the corresponding wavefunction. Furthermore, the completeness equation is satisfied:

$$\int da \, \hat{K}_a^{\lambda\dagger} \hat{K}_a^\lambda = \sqrt{\frac{2\lambda}{\pi}} \int da \, \exp\left[-2\lambda(a-\hat{A})^2\right] = \hat{I} \qquad (16)$$

Finally, while being a mathematical construct, it can be shown by a simplified version of the procedure of Ref. [1] and Appendix A that the measurement operation (15) reflects a physical measurement in the sense that it describes the outcome of a quantum evolution of a system comprizing interacting system and detector.

Coming back to the simultaneous measurement of position and momentum, we have argued above that the measurement operator $\hat{K}_{\alpha_m}$ of Eq. (10), with the measurement result $\alpha_m$ of Eq. (7) representing the obtained position and momentum, corresponds to the strongest simultaneous measurement of these observables. Indeed, the two-observable measurement operator (10) is the closest analog of a projective measurement. Our aim is to construct a systematic protocol for the simultaneous weak measurement of position and momentum, namely a generalization of (14) for such simultaneous measurement. While the availability of such systematic generalization of (10) can be useful in various contexts, our own motivation is to generalize the concept of continuous weak measurement to this case. A consistent description of continuous weak measurement requires a weakness parameter that scales with time. For the measurement of a single observable this is achieved by assuming that $\lambda$ in Eqs. (14) and (16) scales as $\bar{\lambda} dt$ (see, e.g., [17]). An extenstion of (10) to a weak measurement situation, characterized by a suitable weakness parameter $\lambda$, is a prerequisite for an analogous procedure.

We propose such a construction in Section 2 and confirm that it complies with the general requirements of a Kraus operator. In section 3 we show that the proposed mathematical construction can be realized as an actual physical measurement, and also compare the consequences of weak measurement resulting from fuzzy initial detector states and that associated with short duration of the detectors-system interaction. Section 4 summarizes our findings and concludes. In a subsequent publication we will apply the procedures developed here to the description of continuous simultaneous measurement of non-commuting observables and to the analysis of the classical limit of a continously observed system.

## 2. A Krauss operator for modeling weak simultaneous measurement of position and momentum

Here we propose a generalization of the Gaussian Kraus operator (14) to the simultaneous weak measurement of position and momentum. We start with the observation (see Appendix B) that the coherent state representation $\langle \alpha | \Psi \rangle$ of a square-integrable function $\Psi$ is a Gabor transform of this function. In the position representation this takes the form

$$\langle \alpha | \Psi \rangle = \left(\frac{2}{\pi}\right)^{1/4} \int_{-\infty}^{\infty} dx\, e^{-2i\alpha_2 x} e^{-(x-\alpha_1)^2} \langle x | \Psi \rangle \tag{17}$$

Just as the position and momentum representations describe a system state in terms of distributions that fully specify the values of the position or momentum variables, respectively, the coherent states representation uses an (overcomplete) basis of states characterized by minimum uncertainty of these two observables. Being functions of two variables, coherent states reside in the function space $L^2(R^2)$ - square integrable functions of two variables. However, being images of functions in $L^2(R)$, they occupy only a subspace (henceforth denoted G) of the the former. We postulate that, starting from a system in a pure state, a simultaneous measurement of position and momentum leaves the system in another pure state in this subspace. Finally, we note (see Appendix B) that the operator $\hat{P}_G \equiv \pi^{-1} \int d^2\alpha |\alpha\rangle\langle\alpha|$ is a projection operator onto subspace G.

The implications of these statements can be seen by reformulating the consideration of the strongest possible simultaneous measurement of position and momentum. Suppose that such measurement has yielded the values $x_m$ and $p_m$. If position and momentum were independent variables so that the initial state is represented as a function of two variables, $\Psi_B(x,p)$, in $L^2(R^2)$, then following the measurement the wavefunction would collapse to $\Psi_A(x,p) = \Psi_B(x_m, p_m)\delta(x-x_m)\delta(p-p_m)$.[18] This function however is outside subspace G, and we suggest that the strongest possible measurement yields a function in G that is closest to it, namely $\Psi_A = \hat{P}_G \Psi_B(x,p)\delta(x-x_m)\delta(p-p_m)$. This leads to

$$\Psi_A = \pi^{-1}\int dx \int dp |\alpha(x,p)\rangle\langle\alpha(x,p)|\delta(x-x_m)\delta(p-p_m)\rangle \Psi_B(x_m,p_m) \quad (18)$$

while the form $\langle\alpha(x,p)|\delta(x-x_m)\delta(p-p_m)\rangle \Psi_B(x_m,p_m)$ appears unusual, its meaning is clear: We want here a scalar product between $|\alpha(x,p)\rangle$ and a product of eigenvectors of the position and momentum operators that correspond to eigenvalues $x_m$ and $p_m$, respectively. Since the latter is proportional to $\delta(x-x_m)\delta(p-p_m)$ the result of the integration is, up to a constant, $|\alpha(x_m,p_m)\rangle$ in agreement with Eqs. (10)-(11). The normalization constant is not determined by (18) because the projection $\hat{P}_G$ does not necessarily keep normalization, but we know already that $|\alpha(x_m,p_m)\rangle$, Eq. (8), is normalized.

An extension of this procedure may be used to construct an operator for the weak simultaneous measurement of position and momentum that yields $x_m$ and $p_m$ as results. If these variables were independent (with the corresponding operators mutually commuting) and $\Psi_B$ was an eigenstate of both with eigenvalues $x$ and $p$, we could cast the wavefunction following such measurement as a 2-dimensional generalization of Eq. (14), that is

$$\begin{aligned}\Psi_B &\to \left(\frac{2\lambda}{\pi}\right)^{1/2} \exp\left[-\lambda(p_m-p)^2 - \lambda(x_m-x)^2\right]\Psi_B \\ &= \left(\frac{2\lambda}{\pi}\right)^{1/2} \exp\left[-\lambda|\alpha_m-\alpha|^2\right]\Psi_B\end{aligned} \quad (19)$$

where $\alpha = \alpha(x,p); \alpha_m = \alpha(x_m,p_m)$. Obviously, we could use different weakness parameters for the two measurements, but one can always rescale variables to get back the form (19). Projecting onto subspace G of $L_2(R^2)$ we now get

$$\Psi_B = \hat{P}_G \exp\left[-\lambda|\alpha_m - \alpha|^2\right]\Psi_B = N\int d^2\alpha \exp\left[-\lambda|\alpha_m - \alpha|^2\right]|\alpha\rangle\langle\alpha|\Psi_B\rangle, \quad (20)$$

implying the following form of the required operator

$$\hat{K}^\lambda_{\alpha_m} = N\int d^2\alpha \exp\left[-\lambda|\alpha_m - \alpha|^2\right]|\alpha\rangle\langle\alpha| \quad (21)$$

where $N$ is a normalization constant. In appendix C we show that with a proper choice of $N$ this operator satisfies the pre-requisite normalization condition for a Krauss operator

$$\int d^2\alpha_m \hat{K}^\lambda_{\alpha_m} \hat{K}^{\lambda\,\dagger}_{\alpha_m} = 1 \; ; \qquad N = \sqrt{\frac{\lambda(\lambda+2)}{\pi^3}} \quad (22)$$

thus confirming that the set of positive operators $\hat{K}^\lambda_{\alpha_m} \hat{K}^{\lambda\,\dagger}_{\alpha_m}$ is a POVM. We note in passing that such a POVM is not unique. Indeed, the set of properly normalized positive operators $\hat{K}^\lambda_{\alpha_m}$ themselves also constitute a POVM since (using (21)) $\int d^2\alpha_m \hat{K}^\lambda_{\alpha_m} = \pi^2 N/\lambda$. This implies that, in principle, the operators $R^\lambda_{\alpha_m}$ defined by $\left(R^\lambda_{\alpha_m}\right)^2 = K^\lambda_{\alpha_m}$ could also be used as Kraus operators. Both sets are just mathematical constructs of possible measurement processes. Significantly, we show in the following Section that the operators $K^\lambda_{\alpha_m}$ define the outcome of an actual measurement as defined by a procedure analogue to that of Ref. [1].

## 3. Realization of weak simultaneous position-momentum measurement

The Kraus-type operator (10) was shown in Sect. 1 to represent an Arthurs-Kelly (AK) type measurement with the smallest uncertainty $\langle\delta x^2\rangle\langle\delta p^2\rangle$ in the determined position and momentum. Here we show that its extension (21) to weaker measurements with larger uncertainties can be realized in a similar way that differ from the original AK measurement only by the choice of the initial detectors wavefunctions.

To this end, we consider an arbitrary quantum state $|\Psi_B\rangle$ and the resultant state $|\Psi_A\rangle$ defined by the Kraus operator $K^\lambda_{\alpha_m}$ of Eq. (21)

$$|\Psi_A\rangle = K^\lambda_{\alpha_m}|\Psi_B\rangle = N\int d^2\alpha_0 e^{-\lambda|\alpha_0 - \alpha_m|^2}|\alpha_0\rangle\langle\alpha_0|\Psi_B\rangle, \quad (23)$$

where $\alpha = (x_0, p_0)$ and $\alpha_m = (x_m, p_m)$. Our aim is to show that this state is obtained by a modified version of the AK protocol. Obviuisly, the resultant state $|\Psi_A\rangle$ depends parametrically on the result $(x_m, p_m)$ of the simultaneous weak measurement. In the position representation, Eq. (23) takes the form

$$\Psi_A(x; x_m, p_m) = N \int dy \int d^2\alpha \, e^{-\lambda|\alpha_0 - \alpha_m|^2} \langle x|\alpha\rangle\langle\alpha|y\rangle \Psi_B(y), \qquad (24)$$

where we have used $\Psi(x) = \langle x|\Psi\rangle$. Next, using (8) and its complex conjugate, we obtain

$$\Psi_A(x; x_m, p_m)$$
$$= N\left(\frac{2}{\pi}\right)^{1/2} \int dy \int dx_0 \int dp_0 \, e^{-\lambda(x_0 - x_m)^2} e^{-\lambda(p_0 - p_m)^2} e^{-(x - x_0)^2} e^{-(y - x_0)^2} e^{2ip_0(x - y)} \Psi_B(y). \qquad (25)$$

We can bring this expression into a form closer to that established by Arthurs and Kelly[1] for the system-detectors wavefunction at a time $t = K^{-1}$, by introducing the change of variables $\omega = x - y$ such that

$$\Psi_A(x; x_m, p_m)$$
$$= N\left(\frac{2}{\pi}\right)^{1/2} \int d\omega \int dx_0 \int dp_0 \, e^{-\lambda(x_0 - x_m)^2} e^{-\lambda(p_0 - p_m)^2} e^{-(x - x_0)^2} e^{-(x - x_0 - \omega)^2} e^{2ip_0\omega} \Psi_B(x - \omega), \qquad (26)$$

and integrating with respect to $x_0$ and $p_0$. This leads to (using the value of N from Eq. (22))

$$\Psi_A(x; x_m, p_m) = \frac{\sqrt{2}}{\pi} \int d\omega \, e^{-2\lambda/(\lambda+2)(x_m - x + \omega/2)^2} e^{-[(\lambda+2)/2\lambda]\omega^2} e^{2ip_m\omega} \Psi_B(x - \omega). \qquad (27)$$

To connect with the procedure outlined in Appendix A, it is convenient to reverse the change of variables defined in Eq. (5). We get

$$\Psi_A(\bar{x}; \bar{x}_m, \bar{p}_m) = \frac{1}{\pi\sqrt{2b}} \int d\bar{\omega} \, e^{-\lambda/(\lambda+2)\frac{(\bar{x}_m - \bar{x} + \bar{\omega}/2)^2}{b}} e^{-[(\lambda+2)/\lambda]\frac{\bar{\omega}^2}{4b}} e^{i\bar{p}_m\bar{\omega}} \Psi_B(\bar{x} - \bar{\omega}), \qquad (28)$$

which is equivalent to Eq. (40) (for $Kt = 1$) in Appendix A, provided that the following initial detector wavefunctions are used instead of those in Eqs. (42) and (43),

$$D_1(x_1) = \left(\frac{2}{\pi b_1}\right)^{1/4} e^{-\frac{\bar{x}_1^2}{b_1}} \; ; \tag{29}$$

$$D_2(x_2) = \left(\frac{1}{2b_2\pi}\right)^{1/4} e^{-\bar{x}_2^2/4b_2} \Rightarrow \tilde{D}_2(p_2) = \left(\frac{2b_2}{\pi}\right)^{1/4} e^{-b_2\bar{p}_2^2}$$

with

$$b_1 = \frac{(\lambda+2)}{\lambda}b \quad b_2 = \frac{\lambda}{\lambda+2}b \; . \tag{30}$$

Since the measured observables are the position of detector 1 and momentum of detector 2, the relevant wavefunctions are $D_1(x_1)$ and $\tilde{D}_2(p_2)$. In the limit $\lambda \to \infty$ ($b_1, b_2 \to b$) this gives the original measurement scheme as described in Appendix A. For small $\lambda$ the measurement weakness stems from the broadened detector wavefunctions, $D_1(\bar{x}_1) \sim \exp(-\lambda \bar{x}_1^2/2b)$ and $\tilde{D}_2(\bar{p}_2) \sim \exp(-\lambda b \bar{p}_2^2/2)$. This can be seen explicitly by calculating the variances in the measured position and momentum: As in Eq. (6), the joint probability density for measuring $\bar{x}_m$ and $\bar{p}_m$ can be calculated from (28) using $P(\bar{x}_m, \bar{p}_m) = \int_{-\infty}^{\infty} dx |\Psi_A(\bar{x}; \bar{x}_m, \bar{p}_m)|^2$. We find (Appendix D) that the variances in the weakly measured position and momentum are given by

$$\langle \delta \bar{x}_m^2 \rangle = \langle \delta \bar{x}^2 \rangle_B + \frac{b_1}{4} + \frac{b_2}{4} \tag{31}$$

$$\langle \delta \bar{p}_m^2 \rangle = \langle \delta \bar{p}^2 \rangle_B + \frac{1}{4b_1} + \frac{1}{4b_2} \tag{32}$$

which leads to Eq. (13) in the limit $\lambda \to \infty$, while for $\lambda \to 0$ it gives $\langle \delta \bar{x}_m^2 \rangle = \langle \delta \bar{x}^2 \rangle_B + b/2\lambda$ and $\langle \delta \bar{p}_m^2 \rangle = \langle \delta \bar{p}^2 \rangle_B + 1/(2b\lambda)$. We have thus shown that the weak measurement defined by the operator (21), (22) corresponds to an AK measurement protocol with initial detector wavefunctions given by Eqs. (29), (30).

Two points should be made concerning these results:

(a) The different roles of the parametrs $b$ and $\lambda$ should be noticed. $b$ is a squeezing parameters and could be set to 1 by rescaling $x$ and $p$ while $\lambda$ controls the actual broadening of the detectors wavefunctions.

(b) As already noticed in Section I, the choice made by AK to set $\tau \equiv Kt = 1$ implies that the shifts, $\bar{x}_m$ and $\bar{p}_m$, in the position and momentum of detector 1 and 2 respectively correspond to the values of the position and momentum of system itself. The uncertainty in the latter variables, $\bar{x}$ and $\bar{p}$, corresponds to the excess noise, $\langle \delta \bar{x}^2 \rangle = \langle \delta \bar{x}_m^2 \rangle - \langle \delta \bar{x}^2 \rangle_B$ and $\langle \delta \bar{p}^2 \rangle = \langle \delta \bar{p}_m^2 \rangle - \langle \delta \bar{p}^2 \rangle_B$ in the former. In the limit $\lambda \to 0$ this implies the following expressions for the noise in the simultaneously measured system position and momentum

$$\langle \delta \bar{x}^2 \rangle = b/2\lambda \; ; \qquad \langle \delta \bar{p}^2 \rangle = 1/(2b\lambda) \tag{33}$$

To end this discussion, we note that one can also consider the consequence of measurement as expressed by Eq. (40) in the limit $\tau \equiv Kt \ll 1$ in which the term of order $\tau^2$ can be disregarded. We then have (cf. Appendix A, Eq. (49))

$$\Psi_A(\bar{x}; \bar{x}_m, \bar{p}_m, \tau) = \sqrt{\frac{2}{\pi}} e^{-(\bar{x}_m - \bar{x}\tau)^2/b} \int d\bar{p}\, e^{-i\bar{p}\bar{x}} \tilde{\Psi}_B(\bar{p}) e^{-b(\bar{p}_m - \bar{p}\tau)^2} \tag{34}$$

where

$$\Psi_B(x) = \int dp\, e^{-ipx} \tilde{\Psi}_B(p) \tag{35}$$

Eq. (34) shows that upon a simultaneous measurement of position and momentum using a short time ($\tau \to 0$) interaction, the wavefunction is transformed in a way that reflects shifts of the position and momentum of the detectors 1 and 2 by amounts $x\tau$ and $p\tau$, respectively, where x and p are the position and momentum associated with the system. The implications of this observation on the measurement process will be discussed elsewhere.

## 4. Summary and conclusions

We have considered together two fundamental quantum mechanical concepts: the simultaneous observation of non-commuting operators, here focusing on the position and momentum of a quantum particle, and weak measurements in which the system state is weakly perturbed at the cost of yielding only partial information on the system. Obviously, the strongest possible measurement of two non-commuting observables cannot be projective. In the specific

12case of position and momentum such measurement can be realized by the AK protocol under which the measurement is affected by coupling the system to two detectors, one responding to the system position and the other to its momentum. The resulting system state is known to be described by a quasi-projection of the initial state onto a coherent state and the measurement is expressed by the operation of the Kraus operator of Eq. (10) where the complex $\alpha$ expresses the measured values of the position and momentum variables. In this paper we have derived the corresponding weak measurement operator, Eqs. (21), (22), characterized by a weakness parameter $\lambda$ that extrapolate between the strongest possible measurement ($\lambda \to \infty$) and the vanishing-strength measurement, ($\lambda \to 0$). We have further shown that this weak measurement operator correspond to a generalized AK measurement protocol that uses suitable broadened detectors' wavefunctions, Eqs. (29) and (30). The resulting operator that affects weak simultaneous measurement of position and momentum can be used to formulate a processes of continuous weak simultaneous measurement of these observables and to explore the classical limit of such process. These issues will be studied in a subsequent article.

**Appendix A**

Here we reproduce the Arthurs-Kelly (AK) scheme for the most accurate determination of position and momentum of a quantum particles using two independent detectors. Note that the position and momentum variables appearing in this appendix correspond (by the chosen form of Eqs. (42) and (43) below) to the barred variables $\bar{x}$ and $\bar{p}$ of Section 1. In departure from the language, but not the contents, of AK, we set the interaction between our system and the detectors to affect a position shift in detector 1 and a momentum shift in detector 2, and assume that following the system-detectors interaction projective measurements made to determine the position of detector 1 and the momentum of detector 2. Specifically, the initial state of the system plus detectors is taken to be

$$\phi(x, x_1, x_2, t=0) = \Psi_B(x) D_1(x_1) D_2(x_2), \tag{36}$$

where $\Psi_B(x)$ is the wavefunction of the measured system before the measurement, while $D_1(x_1)$ and $D_2(x_2)$ are the detectors wavefunctions, all in the position representation. The interaction between system and detectors is chosen to be

$$H_{int} = K(\hat{p}_1 \hat{x} + \hat{x}_2 \hat{p}), \tag{37}$$

where $\hat{x}$ and $\hat{p}$ are the system operators representing the position and momentum whose measurement is required, $\hat{p}_1$ is the momentum operator of detector 1 and $\hat{x}_2$ is the position operator of detector 2. Following AK, we assume that the interaction dominates the time evolution during the time $t$ on which it operates so that

$$\phi(x, x_1, x_2, t) = \exp(-itH_{int})\phi(x, x_1, x_2, t=0)$$
$$= e^{-Ktx(\partial/\partial x_1)} e^{-Ktx_2(\partial/\partial x)} e^{K^2 t^2 x_2(\partial/\partial x_1)} \phi(x, x_1, x_2, t=0) \quad (38)$$

The last form on the right is obtained by using the Baker-Campell-Hausdorff formula to disentangle the $\hat{x}$ and $\hat{p} = -i\partial/\partial x$ opeators. Using (36) leads to

$$\phi(x, x_1, x_2, t) = \Psi_B(x - x_2 Kt) D_1\left(x_1 - xKt + \frac{1}{2} x_2 K^2 t^2\right) D_2(x_2) \quad (39)$$

or, in the $p_2$ representation for detector 2

$$\phi(x, x_1, p_2, t) = \frac{1}{\sqrt{2\pi}} \int dx_2 \Psi_B(x - x_2 Kt) D_1\left(x_1 - xKt + \frac{1}{2} x_2 K^2 t^2\right) D_2(x_2) e^{ip_2 x_2} \quad (40)$$

Eq. (40) describes the now entangled state of the system and detectors at the end of the period $t$ during which the interaction has been turned on. The actual act of measurement is completed by making projective determinations of the position of detector 1 and momentum of detector 2 immediately following this period. (Note that because of the choice of interaction (37), the results of these measurement directly provide information on the position and momentum of the system-particle). Denoting the results of these measurements $x_m$ and $p_m$, the system wavefunction following the measurement is given by

$$\Psi_A(x; x_m p_m, t) = \phi(x, x_m, p_m, t) \quad (41)$$

while $|\Psi_A(x; x_m p_m, t)|^2$ corresponds to the joint probability distribution to obtain $x_m$ and $p_m$ in the measurement *and* to find the system at x. Next, in analogy to AK, we specify to the following detector wavefunctions

$$D_1(x_1) = \left(\frac{2}{\pi b}\right)^{1/4} e^{-x_1^2/b} \; ; \quad D_2(x_2) = \left(\frac{1}{2b\pi}\right)^{1/4} e^{-x_2^2/4b} \quad (42)$$

and will also make use of the wavefunction of detector 2 in the momentum representation

$$\tilde{D}_2(p_2) = \left(\frac{2b}{\pi}\right)^{1/4} e^{-bp_2^2}. \quad (43)$$

Furthermore, following AK, we choose $t = K^{-1}$ and make a change of variable, $x_2 \to u = x - x_2$. This leads to

$$\Psi_A(x; x_m, p_m, t = K^{-1}) = \frac{1}{\pi\sqrt{2b}} e^{-(x_m-x)^2/2b} e^{ip_m x} \int du \, \Psi_B(u) e^{-(x_m-u)^2/2b} e^{-ip_m u} \quad (44)$$

and to the joint probability ditribution to obtain $x_m$ and $p_m$ as results of the measurement

$$P(x_m, p_m, t = K^{-1}) = \int dx \, |\Psi_A(x; x_m, p_m, t = K^{-1})|^2 = \frac{1}{2\sqrt{\pi^3 b}} \left| \int du \, \Psi_B(u) e^{-\frac{(x_m-u)^2}{2b}} e^{-iup_m} \right|^2 \quad (45)$$

For completeness we also examine another limit of this scheme, whereupon instead of taking $\tau \equiv Kt = 1$ we assume that $\tau \square 1$ so that the term of order $\tau^2$ in Eq. (40) can be disregarded. This leads to

$$\phi(x, x_1, p_2, \tau) = \frac{1}{\sqrt{2\pi}} D_1(x_1 - x\tau) \int dx_2 \, \Psi_B(x - x_2\tau) D_2(x_2) e^{-ip_2 x_2} \quad (46)$$

or, using (41) - (43)

$$\Psi_A(x; x_m, p_m, \tau) = \frac{1}{\pi\sqrt{2b}} e^{-(x_m-x\tau)^2/b} \int du \, \Psi_B(x - \tau u) e^{-u^2/4b} e^{-ip_m u} \quad (47)$$

This can be put in a more suggestive form by expanding $\Psi_B$ in momentum eigenstates according to

$$\Psi_B(x - \tau u) = \int dp \, e^{-ip(x-\tau u)} \tilde{\Psi}_B(p) \quad (48)$$

which leads to

$$\Psi_A(x; x_m, p_m, \tau) = \sqrt{\frac{2}{\pi}} e^{-(x_m-x\tau)^2/b} \int dp \, e^{-ipx} \tilde{\Psi}_B(p) e^{-b(p_m-p\tau)^2} \quad (49)$$

We see that the measurement transforms the wavefunction in a way that reflects shifting the position and momentum of the detectors 1 and 2 by amounts $x\tau$ and $p\tau$ respectively where x and p are the postion and momentum associated with the system-particle.

**Appendix B**

Here we cast transformation to coherent state representation, $\Psi \to \pi^{-1} \int d^2\alpha \, |\alpha\rangle\langle\alpha|\Psi\rangle$ in the language of the Gabor transform.[19] While we use the terms position and momentum for the variables $x$ and $p$, the procedure described below regards just their mutual Fourier-transform

association $\tilde{f}(p) = (2\pi)^{-1/2} \int_{-\infty}^{\infty} dp e^{ipx} f(x)$ that connects between the position and momentum representations of a Hilbert-space vector $|f\rangle$. The Gabor transform $f(x) \to F(\alpha_1, \alpha_2)$ was originally introduced by Gabor[19] for the analysis of minimum uncertainty time-frequency signals. In the present context we define it as

$$F(\alpha_1, \alpha_2) = \left(\frac{2}{\pi}\right)^{1/4} \int_{-\infty}^{\infty} dx e^{-2i\alpha_2 x} e^{-(x-\alpha_1)^2} f(x) \tag{50}$$

with the inverse transform being

$$f(x) = \left(\frac{\pi^5}{2}\right)^{1/4} \int_{-\infty}^{\infty} d\alpha_2 e^{2i\alpha_2 x} \int_{-\infty}^{\infty} d\alpha_1 F(\alpha_1, \alpha_2) \tag{51}$$

In the language of coherent states, the transform (50) is just $\langle \alpha | f \rangle = \int dx \langle \alpha | x \rangle \langle x | f \rangle$ where $\alpha = \alpha_1 + i\alpha_2$ and the function $\langle x | \alpha \rangle = \langle \alpha | x \rangle^*$ is given by Eqs. (7)-(8). (In the corresponding signal analysis literature this function is referred to as a Gabor wavelet). Obviously, merely introducing this language does not by itself constitute a new development, however new insight (applied in Section 2 of the main text) may be obtained from the following observations:

(a) Eqs. (50) and (51) establish a bijective correspondence between the functions $f(x)$ in $L^2(R)$ and the functions of two variables $F(\alpha_1, \alpha_2)$ in space G. This space, henceforth referred to as Gabor space, is an image of the space $L^2(R)$ and as such is a subspace of the space $L^2(R^2)$ - the space of all square integrable functions of 2 variables.

(b) All relationships defined in $L^2(R)$ have their equivalent in G. In particular the norm $\|f\| = \left(\int dx |f(x)|^2 dx\right)^{1/2}$ is equal to $\|F\| = \left(\pi^{-1} \int d^2\alpha |F(\alpha_1, \alpha_2)|^2\right)^{1/2}$, where $\int d^2\alpha = \int d\alpha_1 \int d\alpha_2$, and the scalar product $\langle f | g \rangle = \int dx f^*(x) g(x)$ can be expressed as $\langle F | G \rangle = \pi^{-1} \int d^2\alpha F^*(\alpha_1, \alpha_2) G(\alpha_1, \alpha_2)$. Note that these statements correspond to the identity $\langle f | g \rangle = \pi^{-1} \int d^2\alpha \langle f | \alpha \rangle \langle \alpha | g \rangle$ that is familiar in the coherent states literature.

(c) While the operator $P_G = \pi^{-1}\int d^2\alpha |\alpha\rangle\langle\alpha|$ is often regarded as a unit operator, see Eq. (9), it behaves as such only when it operates on functions in G. When operating on a general function $\Phi(\alpha_1, \alpha_2)$ in $L_2(R^2)$ it projects onto G. By definition the projected function $P_G\Phi$ is that function in subspace G that is closest to $\Phi$, namely $\|\Phi - P_G\Phi\| < \|\Phi - F\|$ where $F$ is any other function in G.

**Appendix C**

Here we show that with the proper normalization $\hat{K}_\alpha^\lambda$, Eq. (21), satisfies the closure equation characteristic of a Kraus operator,

$$\hat{I} = \int d^2\alpha \hat{K}_\alpha^\lambda \hat{K}_\alpha^{\lambda\dagger} \tag{52}$$

$$= |N|^2 \frac{\pi}{2\lambda} \int d^2\alpha' \int d^2\alpha'' e^{-\frac{\lambda}{2}|\alpha'-\alpha''|^2} |\alpha'\rangle\langle\alpha'|\alpha''\rangle\langle\alpha''| \tag{53}$$

To this end, we show that the matrix element $\langle x|...|y\rangle$ of the operator on the right is $C\delta(x-y)$ and determine $N$ so that the constant C is 1. Evaluating this matrix element requires some caution, applying phase factors consistently. Using the standard expression

$$\langle\alpha'|\alpha''\rangle = \exp\left(-\frac{|\alpha'|^2 + |\alpha''|^2}{2} + \alpha'^*\alpha''\right) \tag{54}$$

where $\alpha = (\alpha_1, \alpha_2)$ also requires that Eq. (8) is modified according to[20]

$$\langle x|\alpha\rangle = \left(\frac{2}{\pi}\right)^{1/4} e^{2i\alpha_2(x-(1/2)\alpha_1)} e^{-(x-\alpha_1)^2} \tag{55}$$

Using (54) and (55) and carrying out the straightforward integration over $\alpha'$ and $\alpha''$ we find

$$\delta(X-Y) = |N|^2 \sqrt{\frac{2}{\pi}} \frac{\pi^{7/2}}{\sqrt{2\lambda(\lambda+2)}} \delta(X-Y) \tag{56}$$

which implies the choice of $N$ given in Eq. (22).

**Appendix D**

Here we outline the evaluation of the second moments, Eqs. (31) and (32), of the distribution associated with the weakly measured position and momentum. The starting point is the distribution for these measured variables

$$P(\bar{x}_m, \bar{p}_m) = \int_{-\infty}^{\infty} dx \left| \Psi_A(\bar{x}; \bar{x}_m, \bar{p}_m) \right|^2 \tag{57}$$

Consider first $\Psi_A$ as given by Eq. (28). It follows that

$$\langle \bar{x}_m^2 \rangle = \int d\bar{x} \int d\bar{x}_m \int d\bar{p}_m \, \bar{x}_m^2 \, | \Psi_A(\bar{x}; \bar{x}_m, \bar{p}_m) |^2$$
$$= \int d\bar{x} \int d\bar{x}_m \int d\bar{p}_m \, \bar{x}_m^2 \left| \int dz \, e^{-(\bar{x}_m - \bar{x}/2 - z/2)^2 / b_1} e^{-(\bar{x}-z)^2/4b_2} e^{-i\bar{p}_m z} \Psi(z) \right|^2, \tag{58}$$

where we have defined $z = \bar{x}_m - \bar{\omega}$, by considering first the integral with respect to $\bar{p}_m$, so as to find $\int d\bar{p}_m e^{-i\bar{p}_m(z-z')} = 2\pi \delta(z-z')$, and then integrate with respect to $z'$. Next we compute the resulting integrals with respect to $\bar{x}_m$ and $\bar{x}$, in that order, to obtain

$$\int d\bar{x}_m \bar{x}_m^2 e^{-2(\bar{x}_m - x/2 - z/2)^2/b_1} = \frac{1}{4}\left[(\bar{x}-z)^2 + b_1\right]\sqrt{\frac{\pi b_1}{2}} \tag{59}$$

$$\frac{1}{4}\int d\bar{x}(\bar{x}-z)^2 e^{-(\bar{x}-z)^2/2b_2} = \left(z^2 + \frac{b_2}{4}\right)\sqrt{2\pi b_2}. \tag{60}$$

This leads to the result in Eq. (31) after finding that $\langle \bar{x}_m \rangle = \int dx\, x | \Psi_B(x) |^2$ using a similar strategy. To calculate $\langle \delta \bar{p}_m^2 \rangle$ we first determine $\langle \bar{p}_m^2 \rangle = \int d\bar{x} \int d\bar{x}_m \int d\bar{p}_m \, \bar{p}_m^2 \, | \Psi_A(\bar{x}; \bar{x}_m, \bar{p}_m) |^2$ by integrating first with respect to $\bar{x}_m$ and then with respect to $\bar{x}$ as follows

$$\int d\bar{x}_m e^{-(\bar{x}_m - \bar{x}/2 - z/2)^2/b_1} e^{-(\bar{x}_m - \bar{x}/2 - z'/2)^2/b_1} = \sqrt{\frac{b_1 \pi}{2}} e^{-(z'-z)^2/8b_1}, \tag{61}$$

$$\int d\bar{x} e^{-(\bar{x}-z)^2/4b_2} e^{-(\bar{x}-z')^2/4b_2} = \sqrt{2\pi b_2} e^{-(z'-z)^2/8b_2}. \tag{62}$$

After that, we use the identity

$$-2\pi \frac{\partial^2}{\partial z'^2}\delta(z'-z) = \int d\bar{p}_m \bar{p}_m^2 e^{-i\bar{p}_m(z'-z)}, \tag{63}$$

to evaluate the remaining integrals by parts. A similar calculation yields $\langle \bar{p}_m \rangle = \langle \bar{p} \rangle$, which leads to the final result, Eq. (32).

**Acknowledgements.** The research of AN is supported by the Israel-US Binational Science Foundation, the German Research Foundation (DFG TH 820/11-1), The U.S. National Science Foundation (Grant No. CHE1665291) and the University of Pennsylvania. WB was financially supported by DFG through SFB 767. AN and WB were supported by the Kurt Lion Foundation and an EDEN Project. WB wishes to thank Prof. Adam Bednorz useful discussions. AN and MAO thank Prof. Nimrod Moiseyev and Prof. Shahaf Nitzan for helpful discussions.

[14] J. v. Neumann, *Mathematical Foundations of Quantum Mechanics* (Princeton Univ. Press, Princeton, N.J., 1955),

[15] We use subscripts B and A to denote the wavefunctions before and after measurement, respectively. Barred momentum and position variables are used to denote dimensioned variables similar to those used in AK. In the transformation to dimensionless variables (Eq. (5)) we have kept the same notation for the wavefunction before and after the needed scaling.

[16] K. Kraus, in *States, Effects, and Operations Fundamental Notions of Quantum Theory*, edited by A. Böhm, J. D. Dollard, and W. H. Wootters1983).

[17] A. Bednorz, W. Belzig, and A. Nitzan, *Nonclassical time correlation functions in continuous quantum measurement* New Journal of Physics **14** (2012).

[18] As usual, these forms should be understood as limits of discrete representations obtained by considering a finite discrete lattice in position space when the lattice size and the density of lattice points increase to infinity.

[19] D. Gabor, *Theory of Communication, Part 1* J. Inst. of Elect. Eng. Part III, Radio and Communication **93**, 429 (1946).

[20] It can be shown that the form of Eqs. (54) and (55) are consistent with each other as written. More generally, if we assign a general phase to the coherent state $\alpha = (\operatorname{Re}\alpha \equiv \alpha_1, \operatorname{Im}\alpha \equiv \alpha_2)$ such that $\langle x | \alpha \rangle = (2/\pi)^{1/4} e^{2i\alpha_2(x-k\alpha_1)} e^{-(x-\alpha_1)^2}$, it leads to $\langle \alpha' | \alpha'' \rangle = \int dx \langle \alpha' | x \rangle \langle x | \alpha'' \rangle = \exp\left\{ -\frac{|\alpha'|^2 + |\alpha''|^2}{2} + \alpha_1' \alpha_1'' + \alpha_2' \alpha_2'' - i(\alpha_2' - \alpha_2'')(\alpha_1' + \alpha_1'') + 2ik\left(\alpha_1' \alpha_2'' - \alpha_1'' \alpha_2''\right) \right\}$, which in turn leads to (54) and (55) for the choice $k = 1/2$.